# Secondary Flows and Near-Wall Turbulence in Channel Flow with Longitudinal Ribs

**Ranjan Kushwaha, S. Sarkar*, Gautam Biswas**

Department of Mechanical Engineering, IIT Kanpur, Kanpur-208016, India

ranjankk@iitk.ac.in; *subra@iitk.ac.in; gtm@iitk.ac.in

## ABSTRACT

This study employs the Large Eddy Simulation (LES) to investigate secondary flows and near-wall turbulence induced by two types of surface-mounted longitudinal ribs, namely rectangular and triangular, in a channel flow. The Reynolds number ($Re_\tau$), based on friction velocity and channel height H, is set at 220. The rib aspect ratio W/h, where W and h represent the width and height of the rib, is 2, and the rib spacing, S is 0.6H. The results indicate formation of two counter-rotating vortices between the adjacent ribs for both the cases considered. The roughness function is higher with the rectangular rib as compared to that of the triangular rib. At the location of the mid-plane on the rectangular ribs, the wall shear stress is relatively lower as compared to that of the location of mid-plane between the ribs. Conversely, for the case of triangular rib, the opposite pattern is observed. Normal Reynolds stress exhibits strong anisotropic behaviour near the wall for both the cases, overlapping above 0.4H. Between 0.4H and 0.8H, the variation in normal Reynolds stresses is linear. Near the wall, higher production of turbulent kinetic energy (TKE) and normal Reynolds stresses are observed with the triangular rib as compared to those of the rectangular rib. The ratio of production to dissipation is unity in the log-law region.

## 1. INTRODUCTION

Traditionally it has been believed that surface roughness affects flow behavior only within a thin layer close to the surface, known as the roughness layer. However, recent studies have altered this conjecture, revealing that significant secondary motions can be developed over rough surfaces due to the gradient of Reynolds stress distribution in the spanwise direction. These secondary flows extend their influence on the outer regions of the boundary layer. By adjusting the spacing between roughness elements to be commensurate with the boundary layer thickness, large circulations or secondary flows can be generated. These flows enhance the transport of momentum and energy from the wall and can reduce drag, depending on the geometric parameters of the ribs and their spacing. Therefore, a comprehensive understanding of the flow dynamics over longitudinal ribs is crucial for many important engineering applications [1-3].

## 2. LITERATURE REVIEW AND OBJECTIVE

The present study focuses on the roughness of longitudinal ribs, particularly examining ribs with rectangular cross-sections, a topic investigated by various researchers [4-5] key parameter in characterizing the spanwise heterogeneity of the surface is the spacing between adjacent ribs, which significantly influences the size and strength of the secondary flows. The recent studies investigated the transition and turbulence region influenced by surface roughness [6-7]. As the rib spacing increases, secondary currents expand and eventually fill the space, reaching maximum strength when the rib spacing aligns with the outer length scale of the flow [4]. Beyond this point, further increases in spacing lead to the development of tertiary flows, consisting of pairs of counter-rotating streamwise vortices located in the valleys between ridges [3]. While spanwise spacing has been studied for various ridge shapes [8], other parameters such as ridge width or Reynolds number have predominantly been explored for rectangular ridges. Consequently, there is limited understanding of whether findings from rectangular ribs apply to other ribs shapes. The current study aims to fill this gap by investigating the behavior of secondary currents and comparing average and turbulent quantities for the rectangular and triangular ribs using the Large Eddy Simulation (LES). The research demonstrates a comparative analysis between two types of longitudinal ribs. The study also aims to examine the effects on roughness function, wall shear stress, secondary flow, Reynolds stresses and TKE production.

## 3. FORMULATION AND METHOD

This section discusses computational methodology and the geometry of interest. In LES, the emphasis is on directly calculating the dynamics of large, energy-containing structures that are critical for the transport of momentum and energy. To incorporate the effects of smaller scales, subgrid-scale modeling is utilized. These smaller scales are generally more consistent and isotropic, with lesser dynamic impact. In the governing equations, the effect of these unresolved smaller scales on the resolved scales is represented by a residual stress term. This term is derived from the grid-scale velocity field to close the set of governing equations effectively. The filtering operation is applied to the incompressible Navier–Stokes and continuity equations, resulting in filtered equations of motion that describe the evolution of the significant energy-carrying eddies. These filtered equations are expressed in tensor notation as follows

$$\frac{\partial \tilde{u}_i}{\partial x_i} = 0 \tag{1}$$

$$\frac{\partial \tilde{u}_i}{\partial t} + \frac{\partial \tilde{u}_i \tilde{u}_j}{\partial x_j} = -\frac{\partial \tilde{p}}{\partial x_i} - \frac{\partial \tau_{ij}}{\partial x_j} + \frac{1}{Re_\tau}\frac{\partial^2 \tilde{u}_i}{\partial x_j \partial x_j} \tag{2}$$

In the above equations, the indices $i$ and $j$ (where $i, j$ = 1, 2, 3) correspond to the spatial coordinates $x$ (streamwise), $y$ (cross stream), and $z$ (spanwise). In this formulation, $\tilde{u}_i$ represents the instantaneous filtered velocity in the $i$ direction, $\tilde{p}$ signifies the filtered pressure, and $t$ denotes the time variable. Reynolds number ($Re_\tau$) is defined based on friction velocity ($u_\tau$) and channel height ($H$). The flow variables are nondimensionalized using friction velocity and channel height.

The term $\tau_{ij}$ in Eq. (2) symbolizes the influence of small-scale eddies in the large-scale transport equation. The subgrid-scale model introduced by Smagorinsky [9] relies on the gradient transport hypothesis. This hypothesis establishes a correlation between $\tau_{ij}$ and the large-scale strain rate tensor $\tilde{S}_{ij}$. The model posits that the subgrid-scale stress tensor is proportional to the local strain rate, which is given as,

$$\tau_{ij} - \frac{\delta_{ij}}{3}\tau_{kk} = -2\nu_T \tilde{S}_{ij} \tag{3}$$

With

$$\tilde{S}_{ij} = \frac{1}{2}\left(\frac{\partial \tilde{u}_i}{\partial x_j} + \frac{\partial \tilde{u}_j}{\partial x_i}\right) \tag{4}$$

Here $\nu_T$= eddy viscosity; $\delta_{ij}$ = Kronecker delta.

Lilly [10] proposed the following formula to obtain the eddy viscosity as:

$$\nu_T = (C_s \bar{\Delta})^2 |\bar{S}| \tag{5}$$

Here $|\bar{S}| = \sqrt{\left(2\tilde{S}_{ij}\tilde{S}_{ij}\right)}$ and $C_s$ is the Smagorinsky constant, $\bar{\Delta}$ is the grid filter scale. Germano et al. [11] introduced a dynamic procedure for calculating the eddy viscosity coefficient $C$ at each time step and grid point based on the grid-filtered velocity field and test-filtered velocity field [12]. A detailed description pertaining to dynamic subgrid stress model and the solution algorithm incorporated in this work is available elsewhere [13].

The spatial discretization uses a second order central difference scheme and temporal evolution employs the second order accurate Adams-Bashforth scheme [14]. The Poisson equation for pressure correction is solved using the Bi-CGSTAB method, which is an iterative method to solve the system of linear equations, developed by van der Vorst [15].

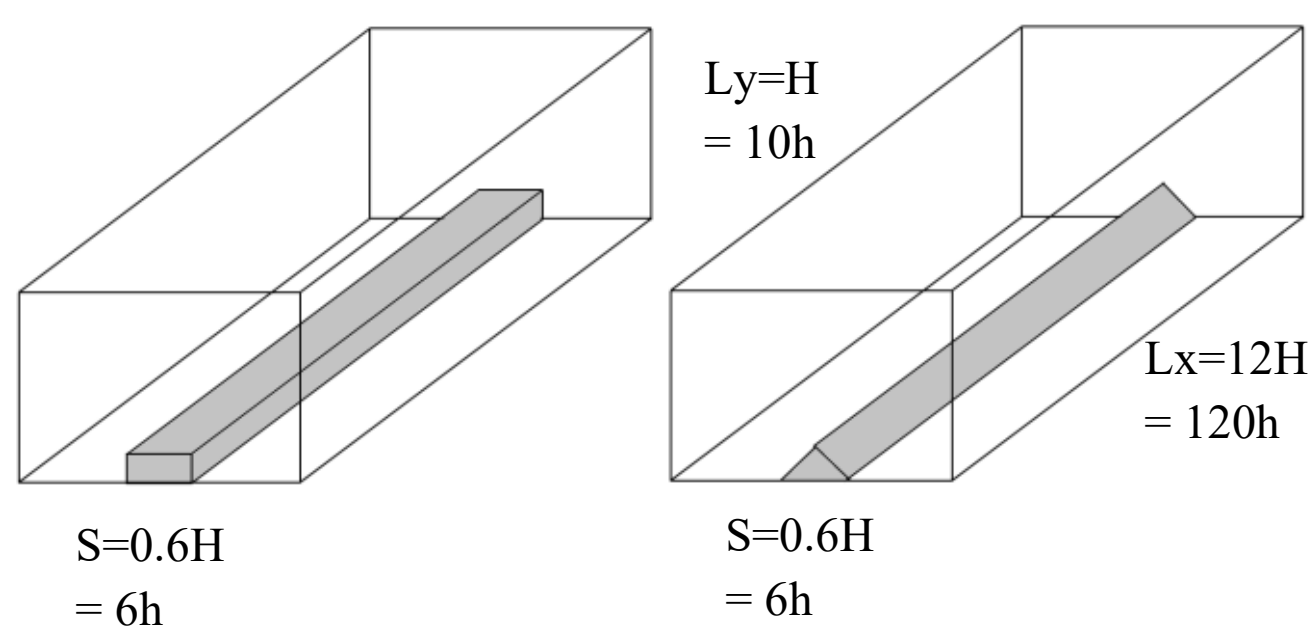

**Figure 1. Computational domain with a single rib; present computation includes six ribs, corresponding to a spanwise domain size of 3.6H.**

The computational domain containing a single rib is depicted in Fig. 1. Our investigations include the configurations with six ribs. The domain size is taken as $12H$ in $x$ direction, $H$ in $y$ direction and $3.6H$ in $z$ direction. Spacing between two ribs is taken as $0.6H$. The width $W$ and Height $h$ of the ribs is taken as $0.2H$ and $0.1H$ respectively. The total number of uniform cells in $x, y$ and $z$ directions is considered as 122, 102 and 362 respectively. The grid resolution, $\delta x^+ = 22.35$ , $\delta z^+$= 2.23 and $\delta y^+$ based on first grid point in the y direction is 1.12, which is sufficient to resolve the laminar sublayer. The grid resolution was carefully selected to accurately capture the turbulent scales while maintaining computational efficiency [16]. Periodic boundary conditions have been applied to ensure continuity in both axial and spanwise directions. The lower boundary enforces a no-slip condition ( $u = v = w = 0$), while the upper boundary adopts a free-slip ($\partial u/\partial y = \partial w/\partial y = v = 0$) condition. The computational parallelization is implemented through Open-MP. The initialization of the velocity field is done by adding the random number to the velocity fields. For the computations, the nondimensional timestep is taken as the order of $10^{-4}$. The Reynolds number $Re_\tau = 220$ has been chosen for this study since it falls in the range used for the studies of near-wall turbulence in the channel flows. Turbulence becomes fully developed when $Re_\tau$ exceeds 150-180 [17], entailing $Re_\tau = 220$ suitable for capturing realistic physics of turbulent flows.

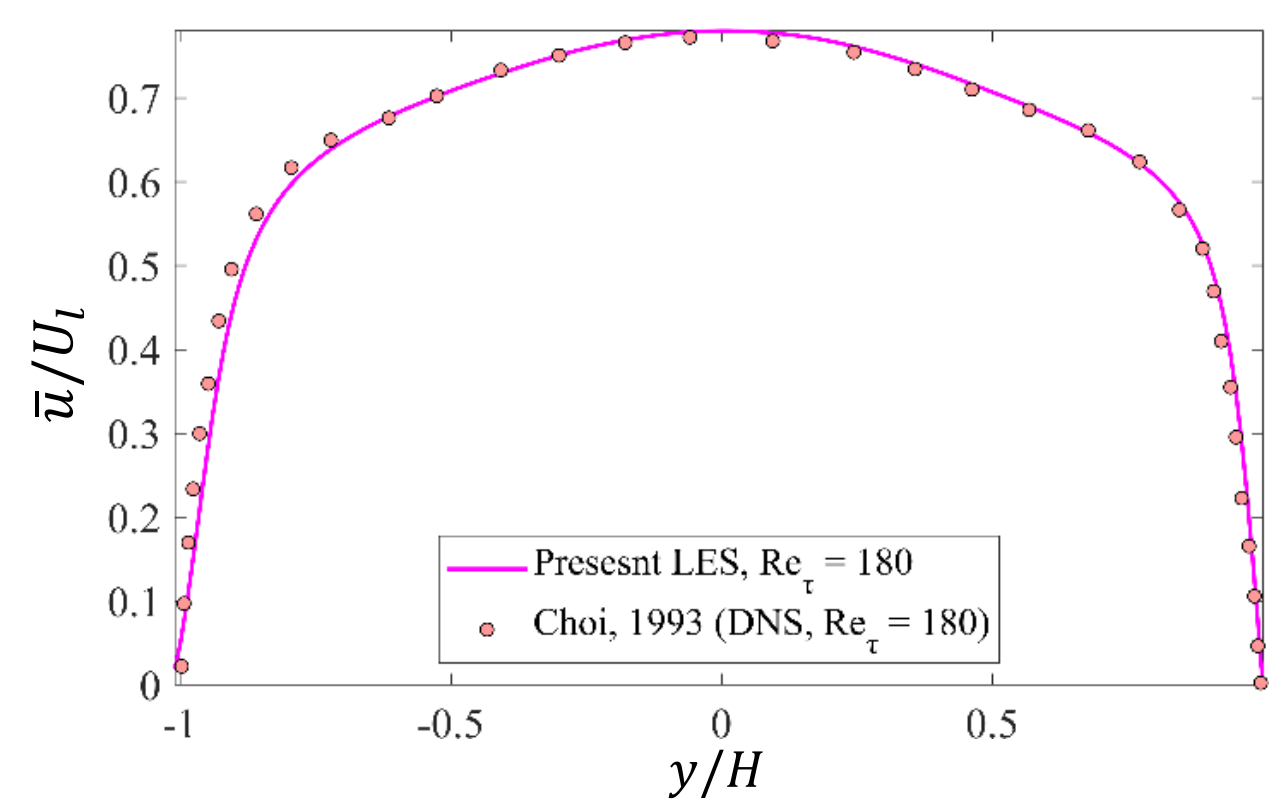

**Figure 2. validation of axial velocity profile**

Our simulation results are validated against the direct numerical simulation (DNS) results of Choi [18] for a channel flow configuration with $Re_\tau$ of 180. In this configuration, the upper wall is a flat plate, while the lower wall features triangular ribs. Periodic boundary conditions have been applied in both axial and spanwise directions. The lower and upper boundary enforces a no-slip condition. Figure 2 shows the validation of the axial velocity profile normalised with centreline velocity of a laminar parabolic profile with the same volume flux, which closely matches the DNS results reported by Choi et al. [18].

## 4. RESULTS AND DISCUSSION

### 4.1 Analysis of averaged velocity

In this subsection, we examine the mean velocity fields. Determining the time-averaged quantities involves applying time integration over an extended duration to obtain converged velocity fields. To establish the axial force balance within the channel, it is essential to equate the shear force with the pressure force. The resulting expressions can be articulated as follows:

Without rib (plane channel):

$$\tau_w \, Sdx = -dp(HS)$$

Then $$\tau_w = \beta H$$

With,

$$\beta = -\frac{dp}{dx}$$

And $$u_\tau = \sqrt{\tau_w} \tag{6}$$

With rectangular rib: -

$$\tau_{wr} \, (S + 2h)dx = -dp(HS - hW)$$

$$\tau_{wr} = \frac{\beta}{(S + 2h)}(HS - hW) \tag{7}$$

And $$u_{\tau r} = \sqrt{\tau_{wr}} \tag{8}$$

With triangular rib:

$$\tau_{wt} \, (S - W + 2\sqrt{2}h)dx = -dp(HS - 0.5hW)$$

$$\tau_{wt} = \frac{\beta}{(S - 2h + 2\sqrt{2}h)}(HS - 0.5hW) \tag{9}$$

And $$u_{\tau t} = \sqrt{\tau_{wt}} \tag{10}$$

In the above derivation for the longitudinal ribs, the wall friction force is calculated using the wall shear stress and wetted area, while the pressure force is determined using the pressure difference and effective area. For the rectangular and triangular rib cases, the shear velocity values, $u_{\tau r}$ and $u_{\tau t}$ are calculated using Eq. 8 and Eq. 10, respectively. The shear velocity value for plane channel $u_\tau$ is calculated using Eq. 6. The ratios of $u_{\tau r}/u_\tau$ and $u_{\tau t}/u_\tau$ are found to be 0.852 and 0.929, respectively. Notably, the ratios $u_{\tau r}/u_\tau$ and $u_{\tau t}/u_\tau$, as determined by computational simulations, are 0.88 and 0.93, closely aligning with the theoretical values. In Fig. 3(a), the averaged velocity, expressed in the wall units, subjected to extrinsic averaging is presented. Extrinsic averaging means that the averaging is done in the axial as well as the spanwise direction across the solid and fluid regions [2]. A log-law profile owing to Moser et al. [19] has been represented by the dashed black line for the plane channel. Surface roughness significantly affects the near-wall velocity profile. This can be seen that due to roughness effect, shift of the buffer layer, causes deviations from the classical log-law profile. Longitudinal rib-surfaces create the secondary flow due to alternation of Reynolds stresses and this alters the velocity distribution. One can see that for a rectangular ribbed channel, extrinsic average axial velocity profile having the roughness function ($\delta\bar{u}^{+}$) value relatively more than the ribbed channel case. The value of $\delta\bar{u}^{+}$ for the triangular rib and rectangular rib case (owing to computations) are found to be 2.1 and 2.8. A kink point can be seen in axial velocity profile (discontinuity) for the computations pertaining to the case of rectangular rib, which was also noted by Castro et al. [2].

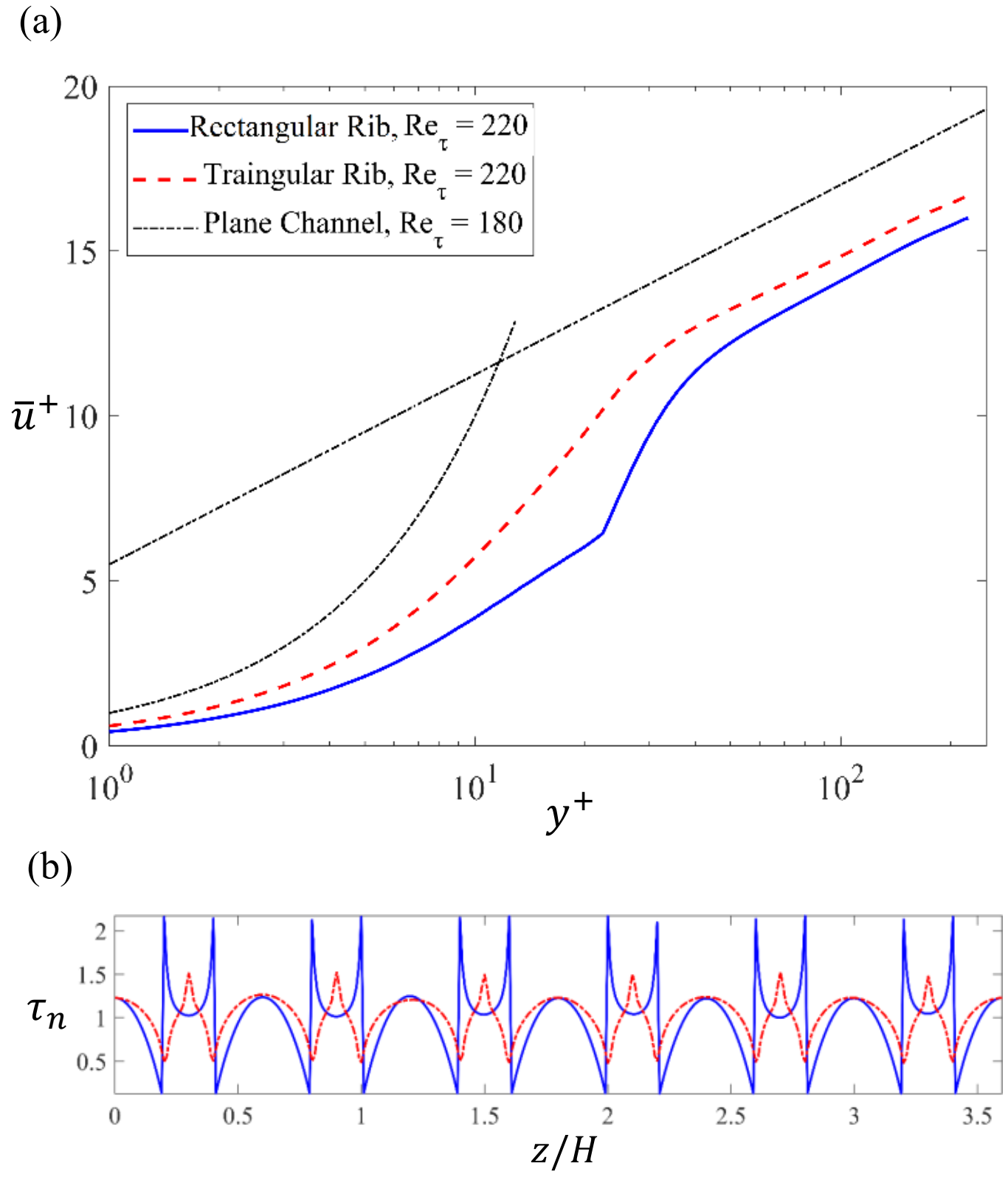

**Figure 3. (a) averaged axial velocity; (b) normalized wall shear stress.**

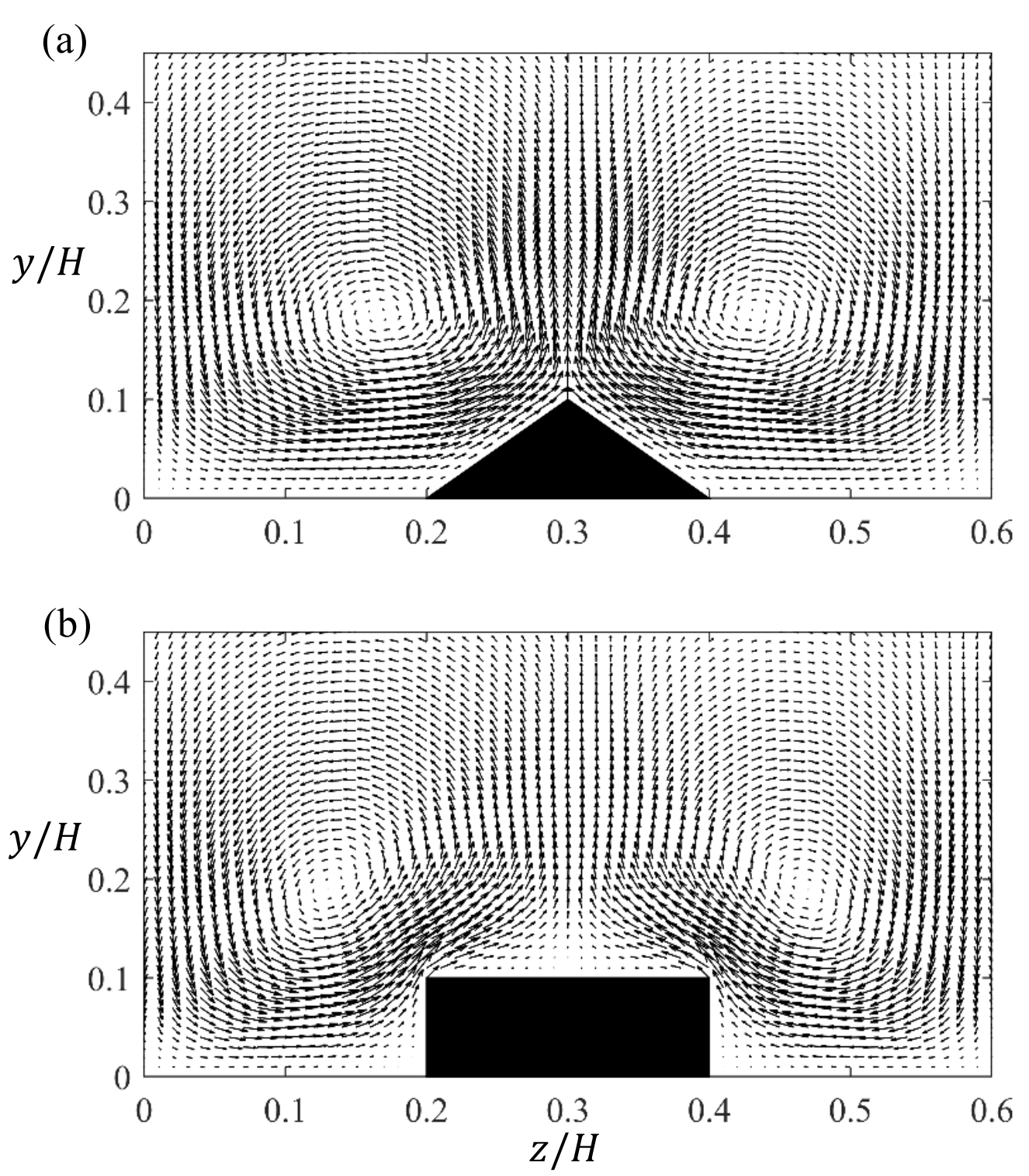

**Figure 4. Secondary flow over (a) triangular rib (b) rectangular rib.**

Figure 3(b) demonstrates the distribution of axial wall shear stress ($\mu\ \partial u/\partial n$ at ($y = 0$)), normalized to achieve an average value of unity [2]. The distribution reveals distinctive patterns. For the case of triangular rib, the wall shear stress is higher compared to the case of rectangular rib at the location of mid-plane of the ribs, as evidenced by the peak values indicated by the red dashed line. In the location of mid-plane between the ribs, the wall shear stress is independent of rib type. For the case of rectangular rib, the wall shear stress is higher at the corners of the ribs, revealed by the peak values as shown by the solid blue line. At the location of mid-plane on the rectangular ribs, the wall shear stress is relatively lower as compared to that of the location of mid-plane between the ribs. Conversely, for the case of triangular rib, the opposite pattern is observed.

In Figs. 4 (a) and (b), secondary flow patterns are depicted using cross-stream and spanwise averaged velocities over the single rib (averaged in streamwise direction due to homogeneity in this direction). It is evident that both types of ribs form a vortex pair. At the mid-plane of the rib, the cross-stream velocity is significantly higher for the case of triangular rib as compared to that of the rectangular rib. Additionally, for the triangular ribs, the dead air zone near the base of the ribs is not visible, whereas for the rectangular ribs, it can be seen at the corners near the base of the rib. The secondary velocity is relatively higher for the triangular rib than that for the rectangular rib. The locations of the left vortex core for the triangular and rectangular ribs are (0.165, 0.184) and (0.130, 0.189), respectively. The vertical location of the vortex core is slightly higher for the case of rectangular rib. The location of the vortex core decides the location of peak value of the production of TKE (discussed in next subsection). For the case of triangular rib, the vortex core forms closer to the mid-plane of the rib.

### 4.2 Analysis of turbulent quantities

In Figs. 5 (a) and (b), the variations of extrinsic averaged Reynolds stresses and production of TKE, for both types of longitudinal ribs, are illustrated. The Reynolds stresses and TKE production owing to the rectangular and triangular ribs, overlap above y/H=0.4, with a linear variation between y/H=0.4 and 0.8. For both the cases, the spanwise normal Reynolds stress surpass the cross-stream normal Reynolds stress. Near the wall, the triangular rib generates relatively higher production of TKE and normal Reynolds stresses as compared to those for the rectangular ribs. As discussed in the previous subsection, the location of the vortex core correlates with the location of the production peak of the TKE. Both rib types exhibit a two-peak pattern in the production plot, one peak is brought about by the bottom wall, and the other peak occurs owing to the roughness. For the case of rectangular rib, the location of the second peak of the TKE production is slightly above than the location of the second peak arising out of the triangular rib. In Fig. 6, the ratio of extrinsic averaged of production to dissipation of (TKE) is illustrated. Within the log-law region, this ratio is close to unity. Close to the wall and above $y/H = 0.3$, the ratios overlap for both types of longitudinal rib shapes.

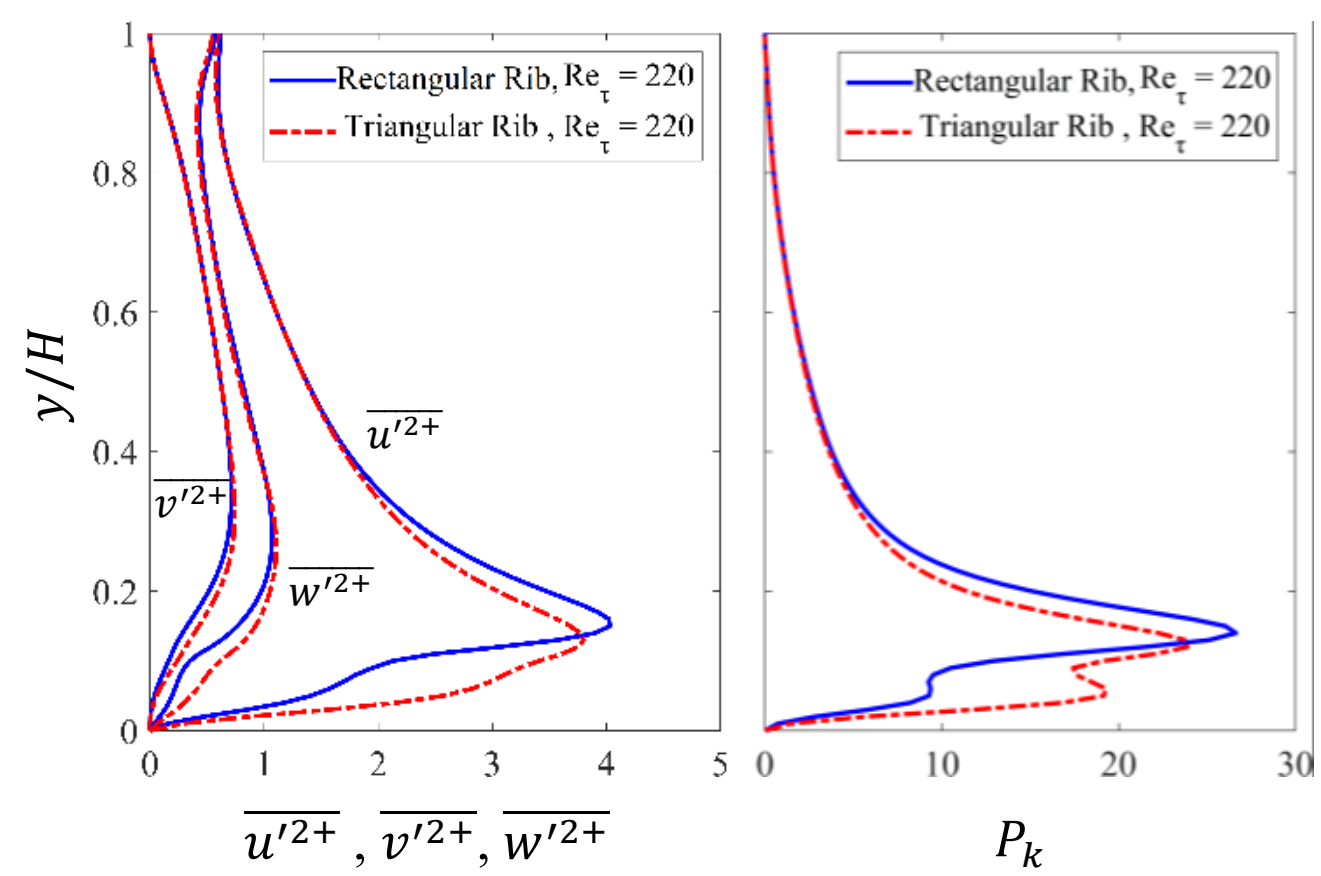

**Figure 5. (a) Reynolds stresses (b) Production of TKE.**

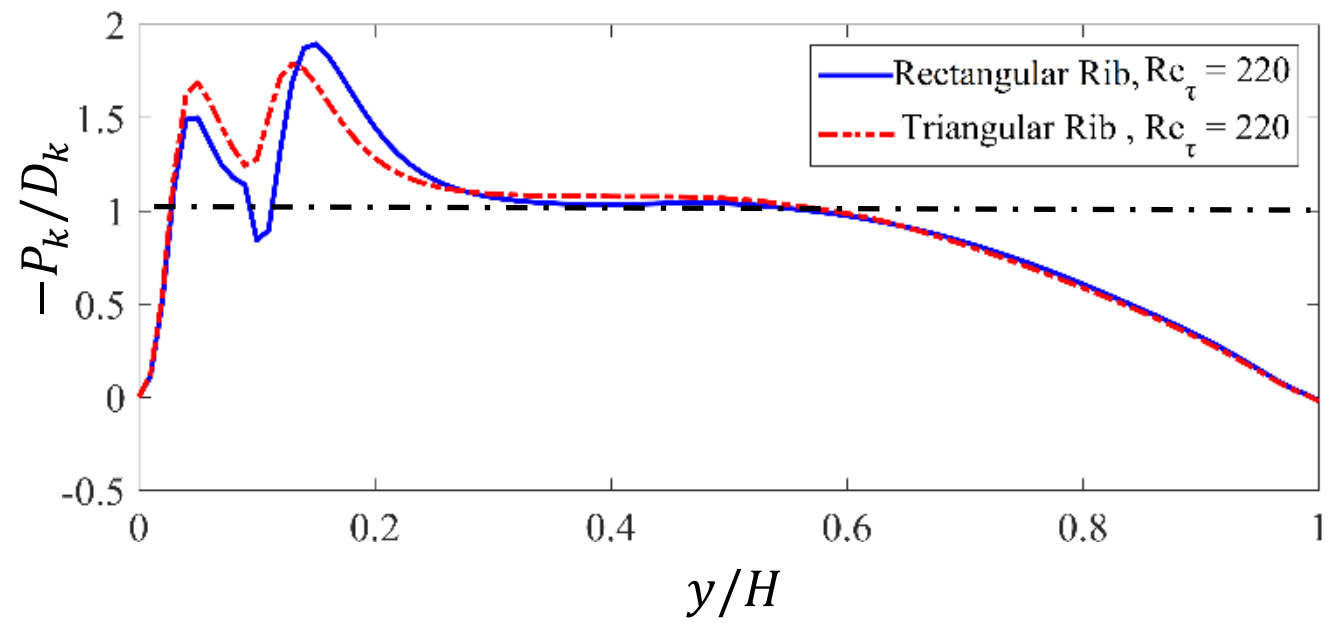

**Figure 6. Ratio of production to dissipation of TKE.**

## 5. CONCLUSIONS

This study deploys LES to understand the comparative behaviour of two types of longitudinal ribs, namely rectangular and triangular, in a channel flow configuration. The roughness function is higher for the rectangular rib compared to that for the triangular rib. At the location of the mid-plane on the rectangular ribs, the wall shear stress is relatively low compared to that at the location of the mid-plane between the ribs. The converse is observed in case of the triangular rib. Both types of ribs form a counterrotating vortex pair, and the vertical location of the vortex core is corelated with the location of production peak of TKE. The DNS results of Castro et al. [2] show such swirling motion (figure 10) on the cross-stream plane. The secondary velocities are relatively higher for the triangular rib. For the triangular ribs, the dead air zone near the base of the ribs is not seen, whereas for the rectangular ribs, it is seen at the corners near the base of the rib. It is also worth mentioning that the results in this work reveal that the cross stream mean velocities are small compared with the mean axial velocity. For both types of longitudinal ribs, the normal Reynolds stresses exhibit strong anisotropic behaviour near the wall, overlapping above 0.4H. Between 0.4H and 0.8H, the variation in normal Reynolds stresses is linear. Near the wall, the triangular rib results in higher production of TKE and normal Reynolds stresses as compared to the rectangular rib. The ratio of production to dissipation is found to be unity in the log-law region.

## ACKNOWLEDGEMENTS

The authors express their sincere gratitude to the high-performance computing facilities (HPC and Param Sanganak) of IIT Kanpur. One of the authors (GB) acknowledges the support of JC Bose National Fellowship of SERB (DST), Government of India.